\documentclass[%
reprint,
superscriptaddress,
amsmath,amssymb,
aps,
onecolumn,
showkeys,
]{revtex4-2}
\usepackage{graphicx}
\usepackage{dcolumn}
\usepackage{bm}
\usepackage{hyperref}
\usepackage{multirow}
\usepackage{adjustbox}
\usepackage{graphics}
\usepackage{subfig}

\begin{document}

\title{Spectroscopy and Radiative Decays of $\Omega_{ccc}$ and $\Omega_{bbb}$ Baryons in a Quark–Diquark Model}

\author{Chaitanya Anil Bokade}
\email{anshul497@gmail.com}
\altaffiliation[ORCID iD: ]{0009-0007-5463-6812}
\author{Bhaghyesh}%
\email{bhaghyesh.mit@manipal.edu; Corresponding author}
\altaffiliation[ORCID iD: ]{0000-0003-3994-9945}
\affiliation{Manipal Institute of Technology\\Manipal Academy of Higher Education, Manipal, 576104, Karnataka, India}

\begin{abstract}
	
We present a comprehensive study of the spectra and radiative decays of the triply heavy baryons $\Omega_{ccc}$ and $\Omega_{bbb}$ within a quark–diquark framework using a screened potential model. The analysis is carried out by solving a relativized Hamiltonian for a two–body bound system: the diquark masses are first determined, after which each baryon is treated as a composite of the diquark and the third quark. Employing the obtained wave functions, we calculate electromagnetic transitions using the E1 and M1 operators. We report mass spectra together with E1/M1 decay widths for radially and orbitally excited states, and systematically compare our results with those from other theoretical approaches.

	
\end{abstract}

\maketitle

\section{\label{sec:Introduction} Introduction}

Heavy baryon physics has evolved into an important area in modern particle physics. This growing interest is driven both by advancements in experimental discoveries \cite{1}, as well as development in theoretical approaches which offer deeper insights into the phenomenology of these states. Quantum chromodynamics (QCD) remains the primary foundation for understanding heavy baryon production and decays. Over the last two decades, both theoretical and experimental research has increased, with substantial contributions from collaborations like as LHCb, CLEO, BABAR, CDF, BELLE, and BESIII. These studies have revealed a wide spectrum of heavy-flavoured hadrons, including heavy mesons, baryons, and exotic multiquark states like tetraquarks and pentaquarks \cite{1}. Baryon states with heavy flavour fall into three categories. The first consists of single heavy baryons, with approximately 70 such states already reported \cite{1}. The second class includes doubly heavy baryons. In 2017, the LHCb Collaboration created a major breakthrough by reporting the observation of the doubly charmed baryon $\Xi^{++}_{cc}(3621)$ in the $\Lambda^{+}_{c}K^{-}\pi^{+}\pi^{+}$ mass spectrum \cite{2}. Its measured mass is about 100 MeV higher than that of the $\Xi^{+}_{cc}(3520)$ state previously reported by the SELEX experiment \cite{3,4}. Subsequent research estimated the lifetime of $\Xi^{++}_{cc}$ \cite{5}. Furthermore, novel decay channels  $\Xi^{++}_{cc} \rightarrow \Xi^{+}_{c}\pi^{+}$ and $\Xi^{++}_{cc} \rightarrow \Xi'^{+}_{c}\pi^{+}$ were identified \cite{6,7}, further improving our understanding of the decay characteristics of these baryons. The third group comprises baryons with three heavy quarks, which are still experimentally elusive. Triply heavy baryons, composed entirely of heavy quarks, provide a clean environment for studying baryonic structure due to the absence of light-quark complications. These systems serve as ideal platforms for refining our understanding of heavy-quark dynamics. However, the necessity to produce three heavy quark–antiquark pairs in a single event makes their experimental detection exceedingly difficult. According to Ref. \cite{8}, triply charmed baryons are unlikely to be detected in $e^{+}e^{-}$ collisions, and the likelihood of $\Omega_{bbb}$ baryons is much lower. The authors estimated by neglecting the charm quark mass in the numerator of the quark propagator and trace evaluations. However, as emphasized in Ref. \cite{9}, this results in significant errors. A more rigorous treatment was undertaken in Ref. \cite{10}, where the full calculation of total and differential cross sections for $e^{+}e^{-} \rightarrow \gamma^{*}/Z^{*} \rightarrow \Omega_{QQQ} + \bar{Q}+ \bar{Q}+ \bar{Q}$ was performed at leading order for different collider energies. Their results reaffirmed that it is hard to observe triply heavy baryons in $e^{+}e^{-}$ collisions. Given their negligible cross sections in $e^{+}e^{-}$ interactions, the search for triply heavy baryons requires the high-energy, high-luminosity environment of hadronic colliders such as the LHC. Initial estimates for their production cross sections at the LHC were provided in Refs. \cite{11,12,13,14}. According to Ref. \cite{15}, with an integrated luminosity of $10\text{ fb}^{-1}$, approximately $10^{4}$ to $10^{5}$ events of triply heavy baryons with $ccc$ and $ccb$ content could be produced at the LHC. Other theoretical work have investigated alternate detection strategies, such as analyses of semi-leptonic and non-leptonic decay modes \cite{16,17,18}, and production in the quark-gluon plasma (QGP) formed in heavy-ion collisions. Several studies \cite{19,20,21,22,23,24} have found that the production of triply heavy baryons in such environments might be much higher than in proton-proton collisions. As a result, identifying triply heavy baryons has become a high-priority objective for future heavy-ion research at the LHC \cite{25}. The potential for possible experimental identification has motivated various kinds of theoretical investigations into the properties of triply heavy baryons, employing methods such as Lattice QCD \cite{26,27,28,29,30,31,32,33}, Non-relativistic Constituent Quark Models (NRCQM) \cite{34,35,36,37,38}, the Diquark Model \cite{39,40}, QCD Sum Rules \cite{41,42,43,44,45}, Faddeev Equations \cite{46,47,48,49}, Regge Theory \cite{50,51}, the Bag Model \cite{52,53}, the Hyper-central Quark Model (HCQM) \cite{54,55,56}, Variational Methods \cite{57,58}, the Relativistic Quark Model (RQM) \cite{59,60}, Non-relativistic QCD (NRQCD) \cite{61}, and the Renormalization Group Procedure for Effective Particles (RGPEP) \cite{62}. One persistent issue in baryon spectroscopy is the so-called "missing resonance" problem, which refers to the discrepancy between the large number of excited states predicted by theoretical model \cite{34,55} and the comparatively fewer states detected in experiments \cite{63}. A recommended approach to address this issue is to characterize baryons as a system consisting of a quark and a diquark. This diquark model is supported by group theoretical reasoning. In the $SU(3)_{colour}$ symmetry, two quarks can combine in the $\bar{3}$ representation forming a diquark that shares the colour configuration of an antiquark. This similarity allows the quark–diquark system to be addressed using techniques analogous to those used for mesons, where a quark interacts with an antiquark \cite{41}. As a result, the effective decrease in internal degrees of freedom yields a sparser excitation spectrum that may better match experimental observations \cite{64}. In baryons with heavy quarks, such as those with charm or bottom content, there is a distinct separation of energy scales, with the quark mass being much bigger than its momentum $(m \gg mv \gg mv^{2})$ \cite{65}. This separation enables a systematic shift from full QCD its effective nonrelativistic equivalents, specifically NRQCD and pNRQCD \cite{65}. Further simplifications, such as ignoring color state transitions between singlet and octet representations, allow pNRQCD to be reduced to a potential model governed by the Schrödinger equation, which effectively describes the dynamics of such systems \cite{56,65}. Studies focusing on heavy quarkonium system \cite{66} have shown that relativistic corrections are essential, especially for states involving charm quarks. To account for these effects, the relativized quark model was developed \cite{67,68}, which maintains a balance between physical intuition and the incorporation of relativistic dynamics. The promising prospects for discovering triply heavy baryons motivate us to extend our previously developed screened potential model \cite{69,70,71} to investigate these states within the quark–diquark framework and to compare the results with those obtained from three-body approaches. In this work, we specifically examine the $\Omega_{ccc}$ and $\Omega_{bbb}$ baryons within the quark–diquark formalism.

This article is organized as follows: Section \ref{sec:Methodology} presents the theoretical framework adopted to model the internal dynamics of these heavy baryons and the computational techniques applied to solve the relativistic Schrodinger equation. Section \ref{sec:Radiative Transitions} explores the electromagnetic transition mechanisms relevant to these states. The results are thoroughly analyzed in Section \ref{sec:Results and Discussion}, where we also compare our findings with existing theoretical predictions. A summary of the key outcomes and concluding remarks are presented in Section \ref{sec:Conclusion}.

\section{\label{sec:Methodology} Methodology}

The mass spectra of the $\Omega_{ccc}$ and $\Omega_{bbb}$ baryons are evaluated within the quark–diquark approximation, where the diquark is modeled as a bound pair of identical-flavor quarks. The analysis is done in two steps: first, the mass of the diquark is calculated. Then the computed mass of diquark is used to model the baryon as a two-body system consisting of the diquark and the third quark. To incorporate relativistic kinematic effects, we employ a relativistic extension of the non-relativistic Hamiltonian, as outlined in \cite{67,68}.

\begin{equation}
	\label{eq:1}
	H = \sqrt{-\nabla_{1}^{2} + m_{1}^{2}}+\sqrt{-\nabla_{2}^{2} + m_{2}^{2}} + V(r) \,.
\end{equation}

\noindent Here $\vec{r} = \vec{x}_{1} - \vec{x}_{2}$ is the relative position vector, where $\vec{x}_{1}$ and $\vec{x}_{2}$ denote the spatial coordinates of the two constituents, which could be either two quarks with corresponding masses $m_{1}$ and $m_{2}$ in a diquark system or a quark and a diquark in a baryonic system. The differential operators $\nabla_{1}^{2}$ and $\nabla_2^{2}$ are the Laplacians acting on $\vec{x}_{1}$ and $\vec{x}_{2}$, respectively. To determine the mass spectrum, the Schrodinger equation is solved as an eigenvalue problem, employing the numerical procedure introduced in Refs. \cite{72,73}. A concise overview of this approach is provided below. The relativistic wave equation associated with the Hamiltonian in Eq. \eqref{eq:1} is given by

\begin{equation}
	\left(\sqrt{-\nabla_{1}^{2} + m_{1}^{2}}+\sqrt{-\nabla_{2}^{2} + m_{2}^{2}} + V(r)\right)\Psi(\vec{r})  = E\Psi(\vec{r}) \label{waveeq} \,.
\end{equation}

\noindent Expanding the wave function in terms of spectral integration,

\begin{equation}
	\Psi(\vec{r}) = \int d^{3}r'\int \frac{d^{3}k}{(2\pi)^{3}}e^{i\vec{k}(\vec{r}-\vec{r}')}\Psi(\vec{r}') \,,
\end{equation}

\noindent we can rewrite Eq.\eqref{waveeq} as

\begin{equation}
	\label{eq:2}
	V(r)\Psi\left(\vec{r}\right)+\int d^{3}r'\frac{d^{3}k}{(2\pi)^{3}}\left(\sqrt{k^{2} + m_{1}^{2}}+\sqrt{k^{2} + m_{2}^{2}}\right)e^{i\vec{k}(\vec{r}-\vec{r}')}\Psi\left(\vec{r}'\right) = E\Psi\left(\vec{r}\right) \,.
\end{equation}

\noindent The exponential factor in the Eq.\eqref{eq:2} can be expanded using spherical harmonics as 

\begin{equation}
	e^{i\vec{k}\cdot\vec{r}} = 4\pi\sum_{nl}Y_{nl}^{*}\left(\hat{k}\right)Y_{nl}(\hat{r})j_{l}(kr)i^{l} \label{expterm} \,,
\end{equation}

\noindent where $j_{l}$ denotes the spherical Bessel function, while $Y_{nl}^{*}(\hat{k})$ and $Y_{nl}(\hat{r})$ represent the spherical harmonics, normalized according to $\int d\Omega Y_{n_{1}l_{1}}(\hat{k})Y_{n_{2}l_{2}}(\hat{r}) = \delta_{n_{1}n_{2}}\delta_{l_{1}l_{2}}$. The unit vectors $\hat{k}$ and $\hat{r}$ refer to the directions of $\vec{k}$ and $\vec{r}$, respectively. By expressing the wave function as a product of a radial part $R_{l}(r)$ and an angular part $Y_{nl}(r)$, and substituting Eq. \eqref{expterm} into Eq. \eqref{eq:2}, the equation simplifies as Refs. \cite{72,73}.

\begin{equation}
	\label{eq:3}
	V(r)u_{l}(r) + \frac{2}{\pi}\int dkk^{2}\int dr'rr'\left(\sqrt{k^{2} + m_{1}^{2}}+\sqrt{k^{2} + m_{2}^{2}}\right)j_{l}(kr)j_{l}(kr')u_{l}(r') = Eu_{l}(r) \,.
\end{equation}

\noindent Here $u_{l}(r)$ denotes the reduced radial wave function, related to the full radial wave function by $(R_{l}(r) =  u_{l}(r)/r)$. In the case of bound states, the wave function exhibits rapid decay with increasing separation $r$ and essentially vanishes at large distances. To describe this localization, a characteristic length scale $L$ is introduced, which effectively confines the bound-state wave function within the interval $0<r<L$. Within this finite region, the reduced radial wave function $u_{l}(r)$ can be expressed in a series expansion of spherical Bessel functions corresponding to the orbital angular momentum quantum number $l$ as 

\begin{equation}
	\label{eq:4}
	u_{l}(r) = \sum_{n=1}^{\infty}c_{n}\frac{a_{n}r}{L}j_{l}\left(\frac{a_{n}r}{L}\right) \,,
\end{equation}

\noindent where $c_{n}$ are the expansion coefficients and $a_{n}$ denote the $n$-th roots of the spherical Bessel function, i.e., $j_{l}(a_{n}) = 0$. For sufficiently large $N$, the expansion in Eq. \eqref{eq:4} can be truncated without significant loss of accuracy. The confinement of the wave function leads to a discretization of momentum, allowing the replacement $a_{n}/L \leftrightarrow k$. Consequently, the integral in Eq. \eqref{eq:3} can be approximated by a discrete sum:
\[
\int dk \rightarrow \sum_{n} \frac{\Delta a_{n}}{L}, \quad \text{where } \Delta a_{n} = a_{n} - a_{n-1}.
\]
Within the confined region $0 < r, r' < L$ and incorporating these substitutions into Eq. \eqref{eq:3} results in a final expression involving the discrete coefficients $c_{n}$ \cite{72,73}

\begin{eqnarray}
	\label{eqfinal}
	Ec_{m} &=&
	\sum_{n=1}^{N}\frac{a_{n}}{N_{m}^{2}a_{m}}\int_{0}^{L}drV(r)r^{2}j_{l}\left(\frac{a_{m}r}{L}\right)j_{l}\left(\frac{a_{n}r}{L}\right)c_{n} \nonumber \\  
	&+&\frac{2}{\pi L^{3}}\Delta a_{m}a_{m}^{2}N_{m}^{2}\left(\sqrt{\left(\frac{a_{m}}{L}\right)^{2} + m_{1}^{2}}+\sqrt{\left(\frac{a_{m}}{L}\right)^{2} + m_{2}^{2}}\right)c_{m} \,,
\end{eqnarray}

\noindent where $N_{m}$ is module of spherical Bessel function:

\begin{equation}
	N_{m}^{2} = \int_{0}^{L}dr'r'^{2}j_{l}\left(\frac{a_{m}r'}{L}\right)^{2} \,.
\end{equation}

\noindent Eq. \eqref{eqfinal} represents an eigenvalue problem in matrix form, which is solved numerically. For sufficiently large values of $L$ and $N$, the solutions become stable and converge reliably \cite{72,73}. The resulting eigenvalues correspond to the spin-averaged masses of the bound states, while the associated eigenvectors represent their wavefunctions. The interaction potential $V(r)$ used for the diquark system is given by \cite{74,75,76}

\begin{equation}
	V(r) = \frac{1}{2}V_{V}(r)+\frac{1}{2}V_{S}(r) + V_{SS}(r) \,,
\end{equation}

\noindent where,

\begin{eqnarray}
	\label{potterms}
	V_{V}(r) &=& -\frac{4}{3}\frac{\alpha_{s}(r)}{r} \,,
	\nonumber \\
	V_{S}(r) &=& \lambda\left(\frac{1 - e^{-\mu r}}{\mu}\right) + V_{0}
	\,,
	\nonumber \\
	V_{SS}(r) &=& \frac{16\pi\alpha_{s}(r)}{9m_{1}m_{2}}\left(\frac{\sigma}{\sqrt{\pi}}\right)^{3}e^{-\sigma^{2}r^{2}}\vec{S}_{1}\cdot \vec{S}_{2} \,.
\end{eqnarray}

\noindent Here $V_{V}(r)$ corresponds to the one-gluon exchange interaction, which is dominant at short-range scales. The confinement potential $V_{S}(r)$ is modified to incorporate color screening effects and governs the long-distance behavior of the system. Additionally, the spin-spin interaction given by $V_{SS}(r)$, which includes a smeared delta function to model the hyperfine splitting between singlet and triplet spin configurations. In Eq. \eqref{potterms}, $\alpha_{s}(r)$ is the running coupling constant, which varies with distance. It can be derived in coordinate space through a Fourier transform of the momentum-space coupling $\alpha_{s}\left(Q^{2}\right)$ \cite{67} and is expressed as 

\begin{equation}
	\alpha_{s}(r) = \sum_{i}\alpha_{i}\frac{2}{\sqrt{\pi}}\int_{0}^{\gamma_{i}r}e^{-x^{2}}dx \,,
\end{equation}

\noindent where $\alpha_{i}'s$ are the free parameters fitted to replicate the short distance behavior of $\alpha_{s}\left(Q^{2}\right)$ predicted by QCD. The numerical values for the parameters are $\alpha_{1}=0.15$, $\alpha_{2}=0.15$, $\alpha_{3}=0.20$, and $\gamma_{1}=1/2$, $\gamma_{2}=\sqrt{10}/2$, $\gamma_{3}=\sqrt{1000}/2$ \cite{73}. Since the baryon and diquark consists of identical heavy quarks, the Pauli exclusion principle must be satisfied for both. Given that the color wavefunction of the baryon is antisymmetric, and the flavor wavefunction is symmetric, implying that the combined spin and spatial parts must be symmetric to maintain the overall antisymmetry of the total wavefunction \cite{40}. For $S$- and $D$-wave diqaurk states, the spatial wavefunction is symmetric, necessitating a symmetric spin configuration, which corresponds to a spin $1$ diquark. In contrast, for $P$-wave diquark states, where the spatial part is antisymmetric, the spin component must also be antisymmetric, resulting in a spin $0$ diquark \cite{40}. Once the diquark masses are determined, the baryon masses are calculated using Eq. \eqref{eqfinal}, with the parameters $m_{1}$ and $m_{2}$ replaced by the diquark mass $m_{d}$ and the mass of the third quark $m_{q}$, respectively. The quark–diquark interaction is described by the potential

\begin{equation}
	V(r) = V_{V}(r) + V_{S}(r) \,.
\end{equation}

\noindent The spin-dependent interactions, included perturbatively to remove the degeneracy among baryon states is expressed as \cite{40,77,78}

\begin{equation}
	\label{14}
	V_{SD}(r) = a\vec{L} \cdot \vec{S}_{d} + b\vec{L} \cdot \vec{S}_{q} + c\left( \frac{3}{r^{2}} (\vec{S}_{d} \cdot \vec{r})(\vec{S}_{q} \cdot \vec{r}) - \vec{S}_{d} \cdot \vec{S}_{q} \right) + d\left(\vec{L}_{d} + \vec{S}_{d}\right) \cdot \vec{S}_{q} \,.
\end{equation}

\noindent Here $\vec{L} = \vec{L}_{d} + \vec{L}_{q}$, where $\vec{L}_{d}$ corresponds to the orbital angular momentum between the quarks inside the diquark, and $\vec{L}_{q}$ refers to the orbital angular momentum between the diquark and the third quark. The spin operators for the diquark and the third quark are denoted by $\vec{S}_{d}$ and $\vec{S}_{q}$, respectively.  It has been emphasized that the $c$ and $b$ quarks may not be sufficiently heavy for a diquark composed of these quarks to be considered as a strictly point-like object \cite{77}. Nevertheless, a point-like diquark is not a necessary requirement for the factorization \cite{77}. Corrections due to finite-size of diquarks can be addressed through the introduction of appropriate form factors \cite{40}, however these form factors introduce additional model parameters. Instead of introducing more parameters, we account for the internal structure of the diquark by explicitly including $\vec{L}_{d}$ and its coupling with  $\vec{L}_{q}$ in spin-dependent interactions. The coefficients $a$, $b$, $c$, and $d$ appearing in Eq. \eqref{14} are given by \cite{79}

\begin{align}
	a &= \frac{1}{2m_{d}^{2}} \left( \frac{V_{V}'(r) - V_{S}'(r)}{r} \right) + \frac{1}{m_{d} m_{q}} \left( \frac{V_{V}'(r)}{r} \right) \,,
	\nonumber \\
	b &= \frac{1}{2m_{q}^{2}} \left( \frac{V_{V}'(r) - V_{S}'(r)}{r} \right) + \frac{1}{m_{d} m_{q}} \left( \frac{V_{V}'(r)}{r} \right) \,,
	\nonumber \\
	c &= \frac{1}{3m_{d} m_{q}} \left( \frac{V_{V}'(r)}{r} - V_{V}''(r) \right) \,,
	\nonumber \\
	d &= \frac{32\pi\alpha_{s}(r)}{9m_{d} m_{q}} \left( \frac{\sigma}{\sqrt{\pi}} \right)^{3} e^{-\sigma^{2} r^{2}} \,.
\end{align}

\noindent There are two distinct coupling schemes used to construct the total angular momentum $\vec{J}$ of the baryon from the spin of the diquark $\vec{S}_{d}$, the spin of the third quark $\vec{S}_{q}$, and the orbital angular momentum $\vec{L}$. The first is the $L$–$S$ coupling scheme, where $\vec{S}_{d}$ and $\vec{S}_{q}$ are first combined to form the total spin $\vec{S}$ of the baryon. This spin is then coupled with $\vec{L}$ to obtain the total angular momentum $\vec{J}$. The corresponding basis states are written as  $|^{2S+1}L_{J}\rangle=|\left[\left(L_{d}L_{q}\right)L\left(S_{d}S_{q}\right)S\right]J^{P}\rangle$. The second approach is the $j$–$j$ coupling scheme, which becomes more appropriate in the heavy quark limit. In this scheme, $\vec{L}$ first couples with $\vec{S}_{d}$ to form an intermediate angular momentum $\vec{J}_{d}$, which then combines with $\vec{S}_{q}$ to produce the total $\vec{J}$. The corresponding basis states are given by $|J^{P},J_{d}\rangle=|\left[\left\{\left(L_{d}L_{q}\right)LS_{d}\right\}J_{d}S_{q}\right]J^{P}\rangle$. The transformation between these two coupling bases is governed by the relation \cite{80}

\begin{equation}
	\label{}
	\left| J^{P}, J_{d} \right\rangle = \sum_{S} (-1)^{S_{d} + S_{q} + L + J} 
	\sqrt{(2J_{d} + 1)(2S + 1)} 
	\begin{Bmatrix}
		L & S_{d} & J_{d} \\
		S_{q} & J & S
	\end{Bmatrix}
	\left| {}^{2S+1}L_{J} \right\rangle \,.
\end{equation}

\noindent We diagonalize the mass matrix constructed in the $L$–$S$ basis to convert it into the $j$–$j$ basis since off-diagonal components arise in the $L$–$S$ coupling basis and the physical states are better characterized in the $j$–$j$ coupling scheme in the heavy quark limit. This diagonalization produces the physical mass spectra and the corresponding wavefunctions. The parameters employed in our model are listed in Table~\ref{tab:1}, and are adopted from our earlier work \cite{69,70,71}. Here $m_{q}$ represent the mass of the charm quark for $\Omega_{ccc}$ and bottom quarks for $\Omega_{bbb}$. The parameters $\sigma$, $\lambda$, and $\mu$ denote the smearing parameter for the spin-spin interaction, the QCD string tension, and the screening parameter, respectively.

\begin{table}
	\caption{Parameters used in our model}
	\label{tab:1}
	\centering
	\begin{tabular*}{\textwidth}{@{\extracolsep{\fill}}ccc}
		\hline
		\noalign{\vskip 2pt}
		Parameters & $\Omega_{ccc}$ & $\Omega_{bbb}$ 
		\\
		\hline
		\noalign{\vskip 2pt}
		$m_{q} (GeV)$ & 1.319 & 4.744 \\
		$\sigma (GeV^{2})$ & 1.281 & 4.967 
		\\
		$\lambda (GeV)$ & 0.297 & 0.240 
		\\
		$\mu (GeV)$ & 0.141 & 0.039  
		\\
		\hline
	\end{tabular*}
\end{table}

\section{\label{sec:Radiative Transitions} Radiative Transition}

Beyond the mass spectra, examining the decay behavior of heavy baryons is essential, particularly for aiding their identification in experimental observations. Due to the limited phase space available for hadronic decays, electromagnetic transitions are expected to significantly contribute to the decay mechanisms of low-lying heavy baryon states. In particular, the lowest-lying states of the $\Omega_{ccc}$ and $\Omega_{bbb}$ baryons, the OZI-allowed two-body strong decay modes are suppressed. As a result, radiative transitions become the dominant decay channels for these states \cite{38}. This makes the study of one-photon decay modes particularly relevant for confirming the existence of excited $\Omega_{ccc}$ and $\Omega_{bbb}$ states in experiments \cite{38}. To investigate these radiative decays, we use the method previously validated in several contexts, including the radiative transitions of doubly heavy baryons within the diquark model, singly heavy baryons under the constituent quark model, and in the study of quarkonia \cite{81,82,83,84}. The EM interaction between quarks and photons at tree level is described by the following coupling Hamiltonian \cite{81,82,83,84}

\begin{equation}
	\label{}
	H_{e} = -\sum_{j}e_{j}\bar{\psi_{j}}\gamma_{\mu}^{j}A^{\mu}(\vec{k},\vec{r}_{j})\psi_{j} \,.
\end{equation}

\noindent Here $\psi_{j}$ denotes the field operator corresponding to the $j$-th quark with coordinate $\vec{r}_{j}$, and $A^{\mu}$ represents the photon field carrying three-momentum $\vec{k}$. Since the hadronic wave functions are derived from a Schrodinger-type framework, we express the electromagnetic quark-photon interaction in a non-relativistic form for consistency. Within the rest frame of the decaying hadron, the electromagnetic interaction Hamiltonian can be expanded non-relativistically, yielding the following approximate expression \cite{81,82,83,84}

\begin{equation}
	\label{19}
	h_{e} \simeq \sum_{j} \left[ e_{j}\, \vec{r}_{j} \cdot \vec{\epsilon} - \frac{e_{j}}{2m_{j}}\, \vec{\sigma}_{j} \cdot (\vec{\epsilon} \times \hat{k}) \right] \phi \,,
\end{equation}

\noindent where $m_{j}$ denotes the constituent mass of the $j$-th quark, $\vec{\sigma}_{j}$ is its Pauli spin vector, and $\vec{r}_{j}$ is its coordinate. The $\vec{\epsilon}$ stands for the polarization of the photon, while $\phi = e^{\pm i\vec{k} \cdot \vec{r}_{j}}$ represents the photon plane wave factor, with the positive and negative signs corresponding to photon emission and absorption, respectively. The first term in Eq. \eqref{19} corresponds to the electric transition, whereas the second term captures the magnetic contribution. An important feature of this electromagnetic transition operator is its ability to incorporate the binding effects among constituent quarks. Additionally, this formalism inherently includes contributions from higher-order electromagnetic multipole transitions \cite{83,84}. The corresponding helicity amplitude $\mathcal{A}$ for the radiative decay process can then be formulated as \cite{81,82,83,84}.

\begin{equation}
	\mathcal{A} = -i\sqrt{\frac{\omega_{\gamma}}{2}}\langle f|h_{e}|i \rangle \,,
\end{equation}

\noindent where $\omega_{\gamma}$ is the photon energy. The partial decay widths of EM transitions are given by \cite{81,82,83,84}

\begin{equation}
	\Gamma = \frac{|\vec{k}|^{2}}{\pi} \frac{2}{2J_{i} + 1} \frac{M_{f}}{M_{i}} \sum_{J_{fz},J_{iz}} |\mathcal{A}|^{2} \,.
\end{equation}

\noindent Here $J_i$ refers to the total angular momentum of the initial baryon, while $J_{iz}$ and $J_{fz}$ correspond to the projections of the total angular momentum along the $z$-axis for the initial and final baryon states, respectively. The baryon masses and wave functions employed in the evaluation of the helicity amplitudes are obtained from our model.

\section{\label{sec:Results and Discussion} Results and Discussion}

\begin{figure}[htbp]
	\centering
	\includegraphics[width=1\linewidth]{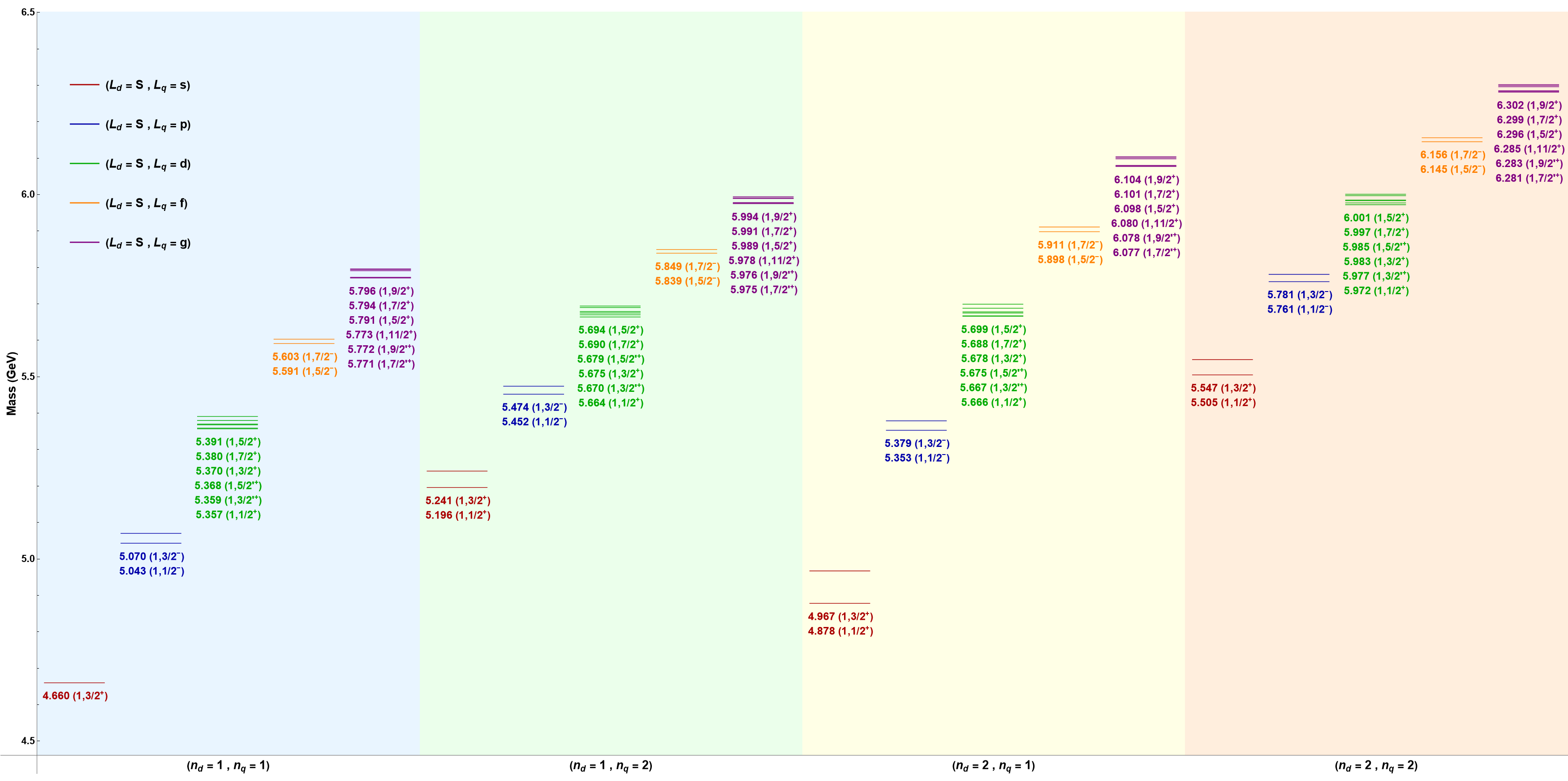}
	\caption{$\Omega_{ccc}$ baryon mass spectra (in GeV) plot.}
	\label{fig:1}
\end{figure}

\begin{figure}[htbp]
	\centering
	\includegraphics[width=1\linewidth]{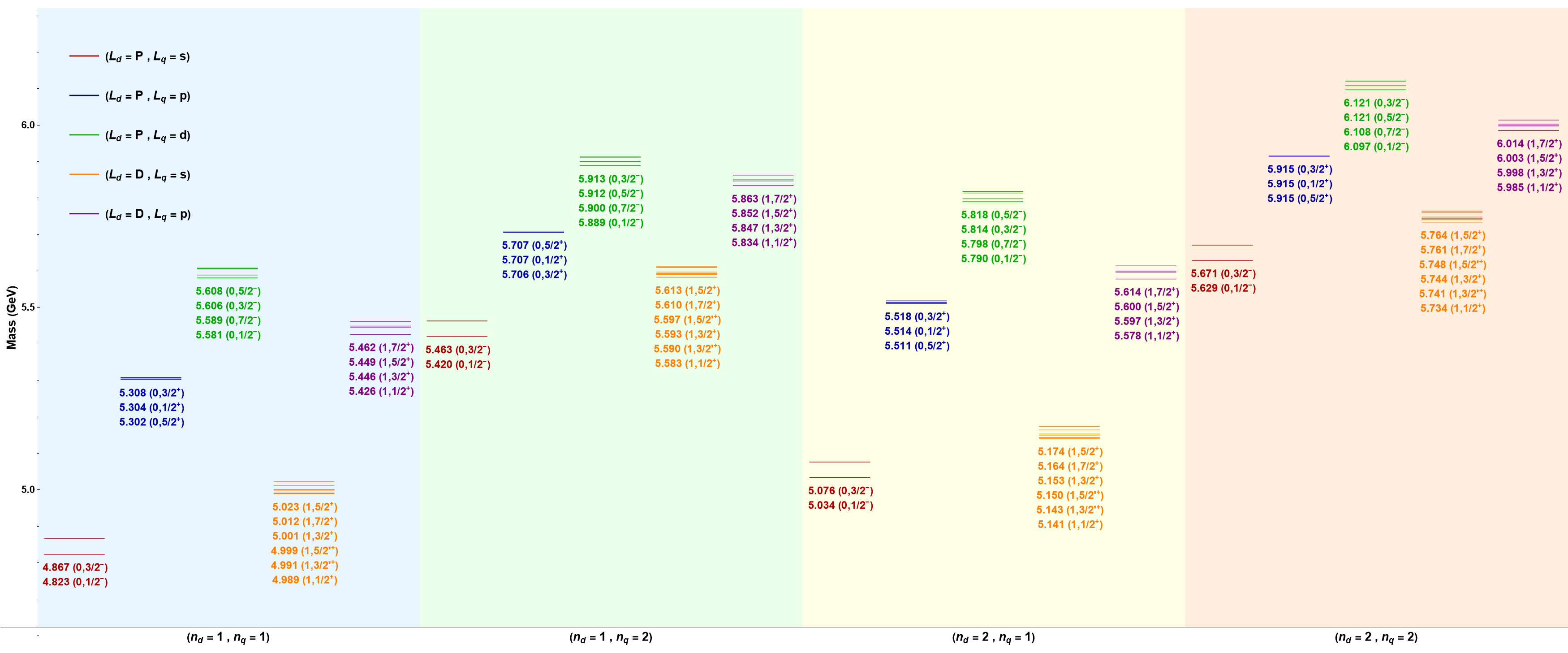}
	\caption{$\Omega_{bbb}$ baryon mass spectra (in GeV) plot.}
	\label{fig:2}
\end{figure}

The $cc$ and $bb$ diquark masses evaluated from our model are listed in Table \ref{tab:2}. These masses are subsequently used as input parameters for evaluating the corresponding baryon masses. The baryon basis states are labeled by the notation $N_{d}L_{d}n_{q}l_{q}$, where $N_{d}$ and $L_{d}$ denote the radial and orbital quantum numbers of the diquark, while $n_{q}$ and $l_{q}$ correspond to those of the third quark in the quark–diquark system. The quantum number for some of these states are summarized in Table \ref{tab:3}. In case of $\Omega_{ccc}$ and $\Omega_{bbb}$ baryons, the presence of three identical quarks requires the total wavefunction to be fully symmetrized. The construction of these symmetrized wavefunctions follows the procedure outlined in Ref. \cite{62}. The resulting symmetrized states of $\Omega_{ccc}$ and $\Omega_{bbb}$ are presented in Table \ref{tab:4}, while the detailed methodology can be found in Ref. \cite{62}. The calculated mass spectra of $\Omega_{ccc}$ and $\Omega_{bbb}$ baryons are presented in Table \ref{tab:5}. For a better visualization, the spectra are also plotted in Figures \ref{fig:1} and \ref{fig:2}, for $\Omega_{ccc}$ and $\Omega_{bbb}$ baryons, respectively. The obtained masses are also compared with results from various theoretical models in Tables \ref{tab:6}, \ref{tab:7}, and \ref{tab:8}.

\begin{table}[!htbp]
	\caption{$S,P$ and $D-$ wave mass spectrum (in MeV) of $cc$ and $bb$ diquarks.}
	\label{tab:2}
	\centering
	\begin{tabular*}{\textwidth}{@{\extracolsep{\fill}}ccc}
		\hline 
		\noalign{\vskip 2pt}
		States & $cc$ & $bb$
		\\ 
		\hline 
		\noalign{\vskip 2pt}
		$1S$ & 3000 & 9595 
		\\
		$2S$ & 3321 & 9905
		\\
		$1P$ & 3233 & 9814 
		\\
		$2P$ & 3451 & 10044
		\\
		$1D$ & 3392 & 9966 
		\\
		$2D$ & 3549 & 10160
		\\
		\hline 
	\end{tabular*}
\end{table}

\begin{table}
	\caption{Quantum numbers of baryon basis states.}
	\label{tab:3}
	\centering
	\begin{tabular*}{\textwidth}{@{\extracolsep{\fill}}ccc}
		\hline 
		\noalign{\vskip 2pt}
		$N_{d}L_{d}n_{q}l_{q}$ & $L \otimes S$ & $J^{P}$
		\\ 
		\hline 
		\noalign{\vskip 2pt}
		$1S1s$ & $0 \otimes \left(\frac{3}{2}\right)$ & $3/2^{+}$ 
		\\
		\noalign{\vskip 2pt}
		$1S1p$ & $1 \otimes \left(\frac{1}{2}\right)_{S}$ & $3/2^{-} \oplus 1/2^{-}$ 
		\\
		\noalign{\vskip 2pt}
		$1P1s$ & $1 \otimes \left(\frac{1}{2}\right)_{A}$ & $3/2^{-} \oplus 1/2^{-}$
		\\
		\noalign{\vskip 2pt}
		$1S1d$ & $2 \otimes \left(\frac{3}{2}\right)$ & $7/2^{+} \oplus...\oplus 1/2^{+}$ 
		\\
		& $2 \otimes \left(\frac{1}{2}\right)_{S}$ & $5/2^{+} \oplus 3/2^{+}$ 
		\\
		\noalign{\vskip 2pt}
		$1D1s$ & $2 \otimes \left(\frac{3}{2}\right)$ & $7/2^{+} \oplus...\oplus 1/2^{+}$ 
		\\
		& $2 \otimes \left(\frac{1}{2}\right)_{S}$ & $5/2^{+} \oplus 3/2^{+}$ 
		\\
		\noalign{\vskip 2pt}
		$1P1p$ & $2 \otimes \left(\frac{1}{2}\right)_{A}$ & $5/2^{+} \oplus 3/2^{+}$ 
		\\
		& $0 \otimes \left(\frac{1}{2}\right)_{A}$ & $1/2^{+}$ 
		\\
		\noalign{\vskip 2pt}
		\hline 
	\end{tabular*}
\end{table}

\begin{table}
	\caption{Symmetrized states of $\Omega_{ccc}$ and $\Omega_{bbb}$ baryons.}
	\label{tab:4}
	\centering
	\begin{tabular*}{\textwidth}{@{\extracolsep{\fill}}ccccc}
		\hline
		\noalign{\vskip 2pt}
		Basis & \multirow{2}{*}{Wavefunction} & \multirow{2}{*}{$J^{P}$} & \multirow{2}{*}{$n^{2S+1}L$} & No of
		\\
		States &  &  &  & States
		\\
		\hline
		\noalign{\vskip 2pt} 
		\multicolumn{5}{c}{Ground State}
		\\
		\hline
		\noalign{\vskip 2pt}
		$1S1s$ & $1S1s\otimes\left(\frac{3}{2}\right)$ & $3/2^{+}$ & $1^{4}S$ & 1
		\\
		\hline 
		\noalign{\vskip 2pt}
		\multicolumn{5}{c}{First Excited States}
		\\
		\hline
		\noalign{\vskip 2pt}
		$1S1p, 1P1s$ & $1P1s\otimes\left(\frac{1}{2}\right)_{A}- 1S1p\otimes\left(\frac{1}{2}\right)_{S} $ & $3/2^{-} \oplus 1/2^{-}$ & $1^{2}P$ & 2
		\\
		\hline
		\noalign{\vskip 2pt}
		\multicolumn{5}{c}{Second Excited States}
		\\
		\hline
		\noalign{\vskip 2pt}
		\multirow{5}{*}{$\begin{array}{c}
				1S2s, 1S1d\\
				2S1s, 1P1p\\
				1D1s
		\end{array}$} & $2S1s_{+}\otimes\left(\frac{3}{2}\right)$ & $3/2^{+}$ & $2^{4}S$ & \multirow{5}{*}{8}
		\\
		\noalign{\vskip 2pt}
		 & $2S1s_{-}\otimes\left(\frac{1}{2}\right)_{S}- 1P1p_{L=0}\otimes\left(\frac{1}{2}\right)_{A} $ & $1/2^{+}$ & $2^{2}S$ &
		\\
		\noalign{\vskip 2pt}
		 & $1D1s_{+}\otimes\left(\frac{3}{2}\right)$ & $7/2^{+}\oplus...\oplus1/2^{+}$ & $1^{4}D$ & 
		\\
		\noalign{\vskip 2pt}
		 & $1D1s_{-}\otimes\left(\frac{1}{2}\right)_{S}- 1P1p_{L=2}\otimes\left(\frac{1}{2}\right)_{A} $ & $5/2^{+}\oplus3/2^{+}$ & $1^{2}D$ &
		\\
		\hline
		\noalign{\vskip 2pt}
		\multicolumn{5}{c}{Third Excited States}
		\\
		\hline
		\noalign{\vskip 2pt}
		\multirow{5}{*}{$\begin{array}{c}
				1S2p,1S1f\\
				2S1p,1P2s\\
				2P1s,1P1d\\
				1D1p,1F1s
			\end{array}$} & \multirow{2}{*}{$\begin{array}{c}
				1P2s\otimes\left(\frac{1}{2}\right)_{A}
				+ 2P1s\otimes\left(\frac{1}{2}\right)_{A}
				+ 1P1d_{L=1}\otimes\left(\frac{1}{2}\right)_{A} \\
				-1S2p\otimes\left(\frac{1}{2}\right)_{S}
				- 2S1p\otimes\left(\frac{1}{2}\right)_{S}
				- 1D1p_{L=1}\otimes\left(\frac{1}{2}\right)_{S}
			\end{array}$} & \multirow{2}{*}{$3/2^{-} \oplus 1/2^{-}$} & \multirow{2}{*}{$2^{2}P$} & \multirow{5}{*}{4} 
		\\ 
		\noalign{\vskip 2pt}
		&  &  &  & 
		\\ 
		\noalign{\vskip 2pt}
		& \multirow{2}{*}{$\begin{array}{c}
				1P1d_{L=3}\otimes\left(\frac{1}{2}\right)_{A} + 1F1s\otimes\left(\frac{1}{2}\right)_{A} \\
				-1D1p_{L=3}\otimes\left(\frac{1}{2}\right)_{S}
				- 1S1f\otimes\left(\frac{1}{2}\right)_{S}
			\end{array}$} & \multirow{2}{*}{$7/2^{-} \oplus 5/2^{-}$} & \multirow{2}{*}{$1^{2}F$} & 
		\\
		\noalign{\vskip 2pt}
		&  &  &  & 
		\\
		\noalign{\vskip 2pt}
		\hline
		\noalign{\vskip 2pt}
		\multicolumn{5}{c}{Fourth Excited States}
		\\
		\hline
		\noalign{\vskip 2pt}
		\multirow{10}{*}{$\begin{array}{c}
				1S3s,1S2d\\
				1S1g,2S2s\\
				2S1d,1P1f\\
				2P1p,1D2s\\
				1D1d,2D1s\\
				1F1p,1G1s\\
			\end{array}$} & $3S1s_{+}\otimes\left(\frac{3}{2}\right) + 2S2s\otimes\left(\frac{3}{2}\right) + 1D1d_{+L=0}\otimes\left(\frac{3}{2}\right)$ & $3/2^{+}$ & $3^{4}S$ & \multirow{10}{*}{14}
		\\
		\noalign{\vskip 2pt}
		& \multirow{2}{*}{$\begin{array}{c}
				3S1s_{-}\otimes\left(\frac{1}{2}\right)_{S} + 2S2s\otimes\left(\frac{1}{2}\right)_{S} \\
				+ 1D1d_{-L=0}\otimes\left(\frac{1}{2}\right)_{S} - 2P1p_{+L=0}\otimes\left(\frac{1}{2}\right)_{A}
			\end{array}$} & \multirow{2}{*}{$1/2^{+}$} & \multirow{2}{*}{$3^{2}S$} &
		\\
		\noalign{\vskip 2pt}
		&  &  &  & 
		\\
		\noalign{\vskip 2pt}
		& $ 2D1s_{+}\otimes\left(\frac{3}{2}\right) + 1D2s_{+}\otimes\left(\frac{3}{2}\right) + 1D1d_{+L=2}\otimes\left(\frac{3}{2}\right) $ & $7/2^{+}\oplus...\oplus1/2^{+}$ & $2^{4}D$ &
		\\
		\noalign{\vskip 2pt}
		& \multirow{2}{*}{$\begin{array}{c}
				2D1s_{-}\otimes\left(\frac{1}{2}\right)_{S} + 1D2s_{-}\otimes\left(\frac{3}{2}\right)_{S} + 1D1d_{-L=2}\otimes\left(\frac{3}{2}\right)_{S} \\
				- 2P1p_{+L=2}\otimes\left(\frac{1}{2}\right)_{A} - 1F1p_{+L=2}\otimes\left(\frac{1}{2}\right)_{A}
			\end{array}$} & \multirow{2}{*}{$5/2^{+}\oplus3/2^{+}$} & \multirow{2}{*}{$2^{2}D$} &
		\\
		\noalign{\vskip 2pt}
		&  &  &  & 
		\\
		\noalign{\vskip 2pt}
		& $ 1G1s_{+}\otimes\left(\frac{3}{2}\right) + 1D1d_{+L=4}\otimes\left(\frac{3}{2}\right)$ & $11/2^{+}\oplus...\oplus5/2^{+}$ & $1^{4}G$ &
		\\
		\noalign{\vskip 2pt}
		& $ 1G1s_{1}\otimes\left(\frac{1}{2}\right)_{S} + 1D1d_{+L=4}\otimes\left(\frac{1}{2}\right)_{S} - 1F1p_{+L=4}\otimes\left(\frac{1}{2}\right)_{A}$ & $9/2^{+}\oplus7/2^{+}$ & $1^{2}G$ &
		\\
		\noalign{\vskip 2pt}
		\hline
		\noalign{\vskip 2pt}
		\multicolumn{5}{c}{Fifth Excited States}
		\\
		\hline
		\noalign{\vskip 2pt}
		\multirow{12}{*}{$\begin{array}{c}
				1S3p,1S2f\\
				1S1h,2S2p\\
				2S1f,3S1p\\
				1P3s,1P2d\\
				1P1g,2P2s\\
				2P1d,3P1s\\
				1D2p,1D1f\\
				2D1p,1F2s\\
				1F1d,2F1s\\
				1G1p,1H1s\\
			\end{array}$}
		& \multirow{4}{*}{$\begin{array}{c}
				3P1s\otimes\left(\frac{1}{2}\right)_{A} + 2P2s\otimes\left(\frac{1}{2}\right)_{A} + 1P3s\otimes\left(\frac{1}{2}\right)_{A} \\
				+ 1P2d_{L=1}\otimes\left(\frac{1}{2}\right)_{A} + 1D2p_{L=1}\otimes\left(\frac{1}{2}\right)_{A} + 1F1d_{L=1}\otimes\left(\frac{1}{2}\right)_{A} \\
				- 1S3p_{L=1}\otimes\left(\frac{1}{2}\right)_{S} - 2S2p_{L=1}\otimes\left(\frac{1}{2}\right)_{S} - 3S1p_{L=1}\otimes\left(\frac{1}{2}\right)_{S} \\ 
				- 2D1p_{L=1}\otimes\left(\frac{1}{2}\right)_{S} - 1D2p_{L=1}\otimes\left(\frac{1}{2}\right)_{S} - 1D1f_{L=1}\otimes\left(\frac{1}{2}\right)_{S}
			\end{array}$} & \multirow{4}{*}{$3/2^{-} \oplus 1/2^{-}$} & \multirow{4}{*}{$3^{2}P$} & \multirow{12}{*}{6}
		\\
		\noalign{\vskip 2pt}
		&  &  &  & 
		\\
		\noalign{\vskip 2pt}
		&  &  &  & 
		\\
		\noalign{\vskip 2pt}
		&  &  &  & 
		\\
		\noalign{\vskip 2pt}
		& \multirow{4}{*}{$\begin{array}{c}
				2F1s\otimes\left(\frac{1}{2}\right)_{A} + 1F2s\otimes\left(\frac{1}{2}\right)_{A} + 1P2d_{L=3}\otimes\left(\frac{1}{2}\right)_{A} \\
				+ 1P1g_{L=3}\otimes\left(\frac{1}{2}\right)_{A} + 2P1d_{L=3}\otimes\left(\frac{1}{2}\right)_{A} + 1F1d_{L=3}\otimes\left(\frac{1}{2}\right)_{A} \\
				- 1S2f\otimes\left(\frac{1}{2}\right)_{S} - 2S1f\otimes\left(\frac{1}{2}\right)_{S} - 2D1p_{L=3}\otimes\left(\frac{1}{2}\right)_{S} \\ 
				- 1G1p_{L=3}\otimes\left(\frac{1}{2}\right)_{S} - 1D2p_{L=3}\otimes\left(\frac{1}{2}\right)_{S} - 1D1f_{L=3}\otimes\left(\frac{1}{2}\right)_{S}
			\end{array}$} & \multirow{4}{*}{$7/2^{-} \oplus 5/2^{-}$} & \multirow{4}{*}{$2^{2}F$} &
		\\
		\noalign{\vskip 2pt}
		&  &  &  & 
		\\
		\noalign{\vskip 2pt}
		&  &  &  & 
		\\
		\noalign{\vskip 2pt}
		&  &  &  & 
		\\
		\noalign{\vskip 2pt}
		& \multirow{2}{*}{$\begin{array}{c}
				1H1s\otimes\left(\frac{1}{2}\right)_{A} + 1P1g_{L=5}\otimes\left(\frac{1}{2}\right)_{A} + 1F1d_{L=5}\otimes\left(\frac{1}{2}\right)_{A} \\
				- 1S1h\otimes\left(\frac{1}{2}\right)_{S} - 1G1p_{L=5}\otimes\left(\frac{1}{2}\right)_{S} - 1D1f_{L=5}\otimes\left(\frac{1}{2}\right)_{S}
			\end{array}$} & \multirow{2}{*}{$11/2^{-} \oplus 9/2^{-}$} & \multirow{2}{*}{$1^{2}H$} &
		\\
		\noalign{\vskip 2pt}
		&  &  &  & 
		\\
		\noalign{\vskip 2pt}
		\hline
	\end{tabular*}
\end{table}

\begin{table}
	\caption{Mass spectrum (in MeV) of $\Omega_{ccc}$ and $\Omega_{bbb}$ baryons.}
	\label{tab:5}
	\centering
	\begin{tabular*}{\textwidth}{@{\extracolsep{\fill}}cccc}
		\hline
		\noalign{\vskip 2pt}
		$J^{P}$ & $n^{2S+1}L$ & $\Omega_{ccc}$ & $\Omega_{bbb}$
		\\
		\hline
		\noalign{\vskip 2pt}
		\multirow{4}{*}{$\frac{1}{2}^{+}$} & $2^{2}S$ & 5171 & 14718
		\\ 
		 & $3^{2}S$ & 5529 & 15074
		\\ 
		 & $1^{4}D$ & 5173 & 14708
		\\ 
		 & $2^{4}D$ & 5587 & 15117
		\\ \noalign{\vskip 2pt}
		\multirow{3}{*}{$\frac{1}{2}^{-}$} & $1^{2}P$ & 4933 & 14499
		\\ 
		& $2^{2}P$ & 5378 & 14927
		\\ 
		& $3^{2}P$ & 5705 & 15258
		\\ \noalign{\vskip 2pt}
		\multirow{7}{*}{$\frac{3}{2}^{+}$} & $1^{4}S$ & 4660 & 14200
		\\ 
		& $2^{4}S$ & 5104 & 14635
		\\ 
		& $3^{4}S$ & 5564 & 15086
		\\ 
		& $1^{4}D$ & 5180 & 14715
		\\
		& $1^{2}D$ & 5244 & 14775
		\\ 
		& $2^{4}D$ & 5594 & 15123
		\\
		& $2^{2}D$ & 5619 & 15144
		\\ \noalign{\vskip 2pt}
		\multirow{3}{*}{$\frac{3}{2}^{-}$} & $1^{2}P$ & 4969 & 14513
		\\ 
		& $2^{2}P$ & 5407 & 14939
		\\ 
		& $3^{2}P$ & 5734 & 15270
		\\ \noalign{\vskip 2pt}
		\multirow{5}{*}{$\frac{5}{2}^{+}$} & $1^{4}D$ & 5099 & 14629
		\\
		& $1^{2}D$ & 5194 & 14733
		\\ 
		& $2^{4}D$ & 5558 & 15081
		\\
		& $2^{2}D$ & 5589 & 15120
		\\
		& $1^{4}G$ & 5632 & 15140
		\\ \noalign{\vskip 2pt}
		\multirow{2}{*}{$\frac{5}{2}^{-}$} & $1^{2}F$ & 5445 & 14965
		\\
		& $2^{2}F$ & 5782 & 15307
		\\ \noalign{\vskip 2pt}
		\multirow{4}{*}{$\frac{7}{2}^{+}$} & $1^{4}D$ & 5196 & 14726
		\\
		& $2^{4}D$ & 5611 & 15134
		\\
		& $1^{4}G$ & 5645 & 15161
		\\
		& $1^{2}G$ & 5669 & 15178
		\\ \noalign{\vskip 2pt}
		\multirow{2}{*}{$\frac{7}{2}^{-}$} & $1^{2}F$ & 5442 & 14967
		\\
		& $2^{2}F$ & 5780 & 15309
		\\
		\hline
	\end{tabular*}
\end{table}

\begin{table}
	\caption{Comparison of the lowest $S-$ wave mass spectra (in MeV) with other theoretical predictions.}
	\label{tab:6}
	\centering
	\begin{tabular*}{\textwidth}{@{\extracolsep{\fill}}ccc}
		\hline
		\noalign{\vskip 2pt}
		Models & $\Omega_{ccc}\left(\frac{3}{2}^{+}\right)$ & $\Omega_{bbb}\left(\frac{3}{2}^{+}\right)$
		\\ \noalign{\vskip 2pt}
		\hline
		\noalign{\vskip 2pt}
		Ours & 4660 & 14200
		\\ 
		Lattice QCD \cite{26} &  & 14371$\pm$4$\pm$11 
		\\
		Lattice QCD \cite{27} & 4761(52)(21)(6) & 
		\\
		Lattice QCD \cite{28} & 4789(6)(21) &  
		\\
		Lattice QCD \cite{29} & 4734(12)(11)(9) & 
		\\
		Lattice QCD \cite{30} & 4763(6) &  
		\\
		Lattice QCD \cite{31} & 4796(8)(18) & 14366(9)(20)
		\\
		Lattice QCD \cite{32} & 4769(6) & 
		\\
		Lattice QCD \cite{33} & 4746(4)(32) & 
		\\
		NRCQM \cite{34} & 4801$\pm$5 & 14373$\pm$25
		\\
		NRCQM \cite{35} & 4965 & 14834
		\\
		NRCQM \cite{36} & 4763 & 14371
		\\
		NRCQM \cite{37} & 4798 & 14396
		\\
		NRCQM \cite{38} & 4828 & 14432
		\\
		Diquark Model \cite{39} & 4760 & 14370
		\\
		Diquark Model \cite{40} & 4712 & 14468
		\\
		QCD Sum Rules \cite{41} & 4670$\pm$150 & 13280$\pm$100
		\\
		QCD Sum Rules \cite{42} & 4720$\pm$120 & 14300$\pm$200
		\\
		QCD Sum Rules \cite{43} & 4990$\pm$140 & 14830$\pm$100
		\\
		QCD Sum Rules \cite{44} & 4810$\pm$100 & 14430$\pm$90
		\\
		QCD Sum Rules \cite{45} & $4530_{-110}^{+260}$ & $14270_{-320}^{+330}$
		\\
		Faddeev Equation \cite{46} & 4900 & 13800
		\\
		Faddeev Equation \cite{47} & 4798.6 & 14244.2
		\\
		Faddeev Equation \cite{48} & 4760 & 14370
		\\
		Faddeev Equation \cite{49} & 4800 & 14370
		\\
		Regge Theory \cite{50} & $4834_{-81}^{+82}$ & 
		\\
		Regge Theory \cite{51} &  & 14788$\pm$80
		\\
		Bag Model \cite{52} & 4790 & 14300
		\\
		Bag Model \cite{53} & 4777 & 14276
		\\
		HCQM \cite{54} & 4812$\pm$85 & 14566$\pm$122
		\\
		HCQM \cite{55} & 4806 & 14496
		\\
		HCQM \cite{56} & 4800 & 14360
		\\
		Variational \cite{57} & 4760$\pm$60 & 14370$\pm$80
		\\
		Variational \cite{58} & 4799 & 14398
		\\
		RQM \cite{59} & 4803 & 14569
		\\
		RQM \cite{60} & 4805 & 14394
		\\
		NRQCD \cite{61} & 4900(250) & 14700(300)
		\\
		RGPEP \cite{62} & 4797 & 14347
		\\
		\hline 
	\end{tabular*}
\end{table}

Figures \ref{fig:1} and \ref{fig:2} presents various radial and orbital excitations spectra of the $\Omega_{ccc}$ and $\Omega_{bbb}$ baryons.  These excitations are of particular interest, as highlighted in Ref. \cite{35}, where the authors suggests that as the quark mass increases, the energy required to excite the diquark decreases. This observation suggests the need to extend the concept of superflavor symmetry to explicitly include internal diquark excitations, as discussed in Ref. \cite{77}.  Table \ref{tab:6} presents our calculated $S$-wave ground state masses for the $\Omega_{ccc}$ and $\Omega_{bbb}$ baryons with $J^P = \frac{3}{2}^{+}$, along with results obtained from other theoretical methods. Our model predicts the $\Omega_{ccc}$ and $\Omega_{bbb}$ ground state masses to be $4660$ MeV and $14200$ MeV, respectively. These values are generally consistent with the range of predictions from other approaches, though they lie at the lower end. Lattice QCD predictions \cite{26,27,28,29,30,31,32,33}, which are based on non-perturbative, first-principles calculations, ranges from $4734-4796$ MeV for $\Omega_{ccc}$ and $14366-14371$ MeV for $\Omega_{bbb}$, with uncertainties spanning tens of MeV. These values are around $70-150$ MeV higher than our predictions,  although the discrepancies are within acceptable theoretical uncertainties. Predictions from NRCQM \cite{34,35,36,37,38} span a broader mass range of $4763-4965$ MeV for $\Omega_{ccc}$ and $14371-14834$ MeV for $\Omega_{bbb}$, which are higher than our estimates, particularly for the $\Omega_{bbb}$ baryon. The discrepancy may be attributed to the non-relativistic three body treatment, in contrast to our two body (quark-diquark) relativized formulation with screening effects. Diquark-based models \cite{39,40} also predict masses higher than our values, with $\Omega_{ccc}$ masses around $4712-4760$ MeV and $\Omega_{bbb}$ about $14370-14468$ MeV. QCD sum rules estimates \cite{41,42,43,44,45} exhibit significant spread and uncertainty, with predictions for $\Omega_{ccc}$ spanning from as low as $4670$ MeV \cite{41} to as high as $4990$ MeV \cite{43}, and for $\Omega_{bbb}$ from $13280$ MeV to $14830$ MeV. This wide variation reflects the model-dependent nature of QCD sum rule approach and sensitivity to threshold parameters. The Faddeev approach \cite{46,47,48,49} also shows wide variation in their predictions with mass values ranging from $4760–4900$ MeV for $\Omega_{ccc}$ and $13800–14370$ MeV for $\Omega_{bbb}$, where our predictions aligns with the lower bound of these estimates. Mass estimates from Regge theory \cite{50,51} and NRQCD \cite{61} are among the highest and the Bag Model also predict mass values higher compared to ours by $\sim$100 MeV for $\Omega_{ccc}$ and $\Omega_{bbb}$. The estimates from HCQM \cite{54,55,56}, Variational \cite{57,58}, RQM \cite{59,60} and RGPEP \cite{62} are silmilar, with mass values around $\sim$4800 MeV for $\Omega_{ccc}$ and a mass range of $14360-14569$ MeV for $\Omega_{bbb}$, which are higher than our prediction. The lower mass in our model reflects the impact of binding dynamics of quark-diquark, relativized kinematics, and the inclusion of screening in the potential. Overall, our predictions for both baryons lie within the general envelope defined by other models.

\begin{table}
	\caption{Comparison of the lowest $P-$ wave mass spectra (in MeV) with other theoretical predictions.}
	\label{tab:7}
	\centering
	\begin{tabular*}{\textwidth}{@{\extracolsep{\fill}}ccccc}
		\hline
		\noalign{\vskip 2pt}
		\multirow{3}{*}{Models} & \multicolumn{2}{c}{$\Omega_{ccc}$} & \multicolumn{2}{c}{$\Omega_{bbb}$}
		\\ 
		\noalign{\vskip 2pt}
		\cline{2-3}
		\cline{4-5} 
		& $\left(\frac{1}{2}^{-}\right)$ & $\left(\frac{3}{2}^{-}\right)$ & $\left(\frac{1}{2}^{-}\right)$ & $\left(\frac{3}{2}^{-}\right)$ 
		\\ 
		\noalign{\vskip 2pt}
		\hline
		\noalign{\vskip 2pt}
		Ours & 4933 & 4969 & 14499 & 14513
		\\
		Lattice QCD \cite{26} &  &  & 14706.3$\pm$9.8$\pm$18.4 & 14714$\pm$9.5$\pm$18.2
		\\
		Lattice QCD \cite{30} & 5120(9) & 5124(13) &  & 
		\\
		NRCQM \cite{35} & 5155 & 5160 & 14975 & 14976
		\\
		NRCQM \cite{37} & 5129 & 5129 & 14688 & 14688
		\\
		NRCQM \cite{38} & 5142 & 5162 & 14773 & 14779
		\\
		Diquark Model \cite{40} & 5010 & 5029 & 14698 & 14702
		\\
		QCD Sum Rules \cite{42} &  & 4900$\pm$100 &  & 14900$\pm$200
		\\
		QCD Sum Rules \cite{43} &  & 5110$\pm$150 &  & 14950$\pm$110
		\\
		Faddeev Equation \cite{48} &  & 5027 &  & 14771
		\\
		Regge Theory \cite{50} &  & $5073_{-107}^{+109}$ &  & 
		\\
		Regge Theory \cite{51} &  &  &  & 15055$\pm$101
		\\
		Bag Model \cite{52} & 5140 &  & 14660 & 
		\\
		HCQM \cite{55} & 5002 & 4982 & 14941 & 14935
		\\
		RQM \cite{60} & 5083 & 5091 & 14682 & 14683
		\\
		RGPEP \cite{62} & 5103 & 5103 & 14645 & 14645
		\\
		\hline 
	\end{tabular*}
\end{table}

\begin{table}
	\caption{Comparison of the lowest $D-$ wave mass spectra (in MeV) with other theoretical predictions.}
	\label{tab:8}
	\centering
	\begin{tabular*}{\textwidth}{@{\extracolsep{\fill}}cccccccccccc}
		\hline
		\noalign{\vskip 2pt}
		\multicolumn{2}{c}{\multirow{2}{*}{States}} & \multirow{2}{*}{Ours} & Lattice & Lattice & NRCQM & NRCQM  & NRCQM & Diquark & HCQM & RQM & RGPEP 
		\\
		&  &  & QCD \cite{26} & QCD \cite{30} &  \cite{35} &  \cite{37} &  \cite{38} & Model \cite{40} &  \cite{55} &  \cite{60} & \cite{62} 
		\\
		\noalign{\vskip 2pt}
		\hline
		\noalign{\vskip 2pt}
		\multirow{6}{*}{$\Omega_{ccc}$} & $\left(\frac{1}{2}^{+}\right)$ & 5173 &  & 5399(13) & 5325 & 5376 & 5352 & 5278 & 5473 & 5313 &  
		\\
		& $\left(\frac{3}{2}^{+}\right)$ & 5180 &  & 5430(13) & 5313 & 5376 & 5368 & 5267 & 5448 & 5330 & 5358 
		\\
		& $\left(\frac{3'}{2}^{+}\right)$& 5244 &  & 5465(13) &  &  & 5412 & 5277 & 5436 &  & 
		\\
		& $\left(\frac{5}{2}^{+}\right)$ & 5099 &  & 5406(15) & 5329 & 5376 & 5392 & 5278 & 5416 & 5329 & 5358 
		\\
		& $\left(\frac{5'}{2}^{+}\right)$& 5194 &  & 5464(15) & 5343 &  & 5433 & 5290 & 5404 &  & 
		\\
		& $\left(\frac{7}{2}^{+}\right)$ & 5196 &  & 5397(49) & 5331 & 5376 & 5418 & 5291 & 5375 & 5353 & 5358 
		\\
		\hline
		\noalign{\vskip 2pt}
		\multirow{6}{*}{$\Omega_{bbb}$} & 
		$\left(\frac{1}{2}^{+}\right)$ & 14708 &  &  & 15102 & 14894 & 14971 & 14912 & 15306 & 14873 &  
		\\
		& $\left(\frac{3}{2}^{+}\right)$ & 14715 & 14953$\pm$17$\pm$24 &  & 15089 & 14894 & 14975 & 14893 & 15300 & 14900 & 14896 
		\\
		& $\left(\frac{3'}{2}^{+}\right)$& 14775 & 14958$\pm$17$\pm$23 &  &  &  & 15016 & 14905 & 15298 &  & 
		\\
		& $\left(\frac{5}{2}^{+}\right)$ & 14629 & 15005$\pm$18$\pm$24 &  & 15109 & 14894 & 14981 & 14895 & 15293 & 14896 & 14896 
		\\
		& $\left(\frac{5'}{2}^{+}\right)$& 14733 & 14964$\pm$17$\pm$23 &  & 15109 &  & 15022 & 14907 & 15291 &  & 
		\\
		& $\left(\frac{7}{2}^{+}\right)$ & 14726 & 14969$\pm$16$\pm$23 &  & 15101 & 14894 & 14988 & 14909 & 15286 & 14904 & 14896 
		\\
		\hline
	\end{tabular*}
\end{table}

Table \ref{tab:7} presents the predicted $P$-wave mass spectra of the $\Omega_{ccc}$ and $\Omega_{bbb}$ baryons for the $J^{P} = \frac{1}{2}^{-}$ and $\frac{3}{2}^{-}$ states. For the $\Omega_{ccc}$ baryon, our model predicts masses of $4933$ MeV and $4969$ MeV for the $\frac{1}{2}^{-}$ and $\frac{3}{2}^{-}$ states, respectively. In the case of the $\Omega_{bbb}$ baryon, the corresponding $\frac{1}{2}^{-}$ and $\frac{3}{2}^{-}$ states are predicted at $14499$ MeV and $14513$ MeV, respectively. The lattice QCD predictions \cite{26,30} for $\Omega_{ccc}$ and $\Omega_{bbb}$ baryons that are approximately $100-200$ MeV higher than our model. NRCQM  estimates \cite{35,37,38} predict comparatively larger mass values, with the $\Omega_{ccc}$ states lying in the range of $5129-5162$ MeV and the $\Omega_{bbb}$ states between $14688-14976$ MeV, which are among the highest values. The diquark model \cite{40}, prediction is moderately higher for the $\Omega_{ccc}$ baryon, while for the $\Omega_{bbb}$ baryon the estimated masses exceed the present results by nearly $200$ MeV. The QCD sum rules estimates \cite{42,43} exhibit wider uncertainties, ranging from $4900-5110$ MeV for $\Omega_{ccc}$ and $14900-14950$ MeV for $\Omega_{bbb}$. Regge theory \cite{50} and RQM \cite{60} estimates for $\Omega_{ccc}$ are consistent with our results. However, for $\Omega_{bbb}$, the estimates from Regge theory \cite{51} and HCQM \cite{55} are higher than our predictions. The Bag model \cite{52} and RGPEP \cite{62} also predict comparatively larger masses for both $\Omega_{ccc}$ and $\Omega_{bbb}$ compared to our values. Overall, our model mass estimates are on the lower side of the spectrum for first excited states. Table \ref{tab:8}  summarizes the predicted $D$-wave mass spectra of the $\Omega_{ccc}$ and $\Omega_{bbb}$ baryons across all possible $J^{P}$ spin-parity states. For the $\Omega_{ccc}$ baryon, our model predicts masses in the range of $5099-5244$ MeV. In the case of the $\Omega_{bbb}$ baryon, the masses lie between $14629-14775$ MeV. When compared to Lattice QCD results \cite{26,30}, our predictions for $\Omega_{ccc}$ and $\Omega_{bbb}$ are typically lower by about $\sim 200$ MeV. NRCQM predictions \cite{35,37,38} provide a wide spread from $5313-5433$ MeV for $\Omega_{ccc}$ and $14894-15109$ MeV for $\Omega_{bbb}$. These are larger than the masses obtained in the present model, with differences reaching nearly $\sim 250$ MeV. The diquark model \cite{40} estimates are more consistent with our predictions for $\Omega_{ccc}$ but are significantly higher for the $\Omega_{bbb}$ baryon. The HCQM results \cite{55} provide the largest mass estimates among the considered theoretical approaches for both baryons. The RQM \cite{60} and RGPEP \cite{62} predictions are also higher for both $\Omega_{ccc}$ and $\Omega_{bbb}$ baryons. 

Overall, the results indicate that our model mass predictions are on the lower side for both   $\Omega_{ccc}$ and $\Omega_{bbb}$ baryons. Moreover the spectra of the $\Omega_{ccc}$ and $\Omega_{bbb}$ baryons exhibit identical energy level structures, which also has been reported in the non-relativistic three-body quark model based on the Gaussian expansion method \cite{85}. For the positive-parity states with $J^{P} = 1/2^{+}$, Ref. \cite{85} predicts two degenerate $S-D$ wave pairs. A similar feature is also observed in our calculations, where the $2^{2}S_{1/2^{+}}$ and $1^{4}D_{1/2^{+}}$ states have nearly equal masses. In the case of $J^{P}=3/2^{+}$ states, Ref. \cite{85} predicts a $1S$ ground state followed by $2S-3S$ state and two $1D-2D$ pairs. This structure closely resembles the spectra obtained in our model, where the $1^{4}S_{3/2^{+}}$ ground state is followed by the excited $2^{4}S_{3/2^{+}}$ state and the two $D-$ wave states $1^{4}D_{3/2^{+}}$ and $1^{2}D_{3/2^{+}}$. For the $J^{P}=5/2^{+}$, Ref. \cite{85} predicts a $1D-2D$ pair, while for the $J^{P}=7/2^{+}$ states a single $D$-wave state with spin $3/2$, consistent with the spectra obtained in our calculations. The agreement also extends to the negative-parity, where Ref. \cite{85} predicts distinct $P$ and $F-$ wave states, which is again consistent with the spectra obtained in the present quark–diquark framework. Ref. \cite{32} reports a negative electric quadrupole moment for the $\Omega_{ccc}$ baryon, suggesting an oblate charge distribution. This characteristic feature aligns well with the quark–diquark interpretation, in which the intrinsic spin of the diquark and non-spherical quark–diquark correlations can generate an oblate intrinsic charge distribution. A covariant spectator quark–diquark model \cite{86} has shown that non-spherical quark–diquark configurations can generate nonzero quadrupole moments associated with oblate deformation. The observed discrepancies in mass estimates among various theoretical frameworks highlight the persistent difficulties in accurately assessing the spectra of triply heavy baryons.  Even minor differences in how inter-quark dynamics are modeled can lead to significant differences in the predicted mass values. Our model predictions for the masses of $\Omega_{ccc}$ and $\Omega_{bbb}$ align with the broad range reported in the literature and also address some of the limitations as discussed above.

\begin{table}
	\caption{$S$ and $P-$ wave states radiative decays.}
	\label{tab:9}
	\centering
	\begin{tabular*}{\textwidth}{@{\extracolsep{\fill}}cccccc}
		\hline
		\noalign{\vskip 2pt}
		Initial & Final & \multicolumn{2}{c}{$\Omega_{ccc}$} & \multicolumn{2}{c}{$\Omega_{bbb}$}
		\\
		\noalign{\vskip 2pt}
		\cline{3-4}
		\cline{5-6}
		\noalign{\vskip 2pt}
		State & State & $E_{\gamma}$(MeV) & $\Gamma$(KeV) & $E_{\gamma}$(MeV) & $\Gamma$(eV) 
		\\
		\hline
		\noalign{\vskip 2pt}
		$3^{4}S_{3/2^{+}}$ & $2^{2}S_{1/2^{+}}$ & 379.74 & 36.636 & 362.64 & $1.067 \times 10^{3}$
		\\
		 & $3^{2}S_{1/2^{+}}$ & 34.95 & 0.055 & 11.46 & 0.046
		\\
		\hline
		\noalign{\vskip 2pt}
		$1^{2}P_{3/2^{-}}$ & $1^{2}P_{1/2^{-}}$ & 35.29 & 0.007 & 13.39 & 0.010
		\\
		\hline
		\noalign{\vskip 2pt}
		$2^{2}P_{1/2^{-}}$ & $1^{2}P_{3/2^{-}}$ & 393.55 & 5.402 & 408.26 & 134.360
		\\
		\hline
		\noalign{\vskip 2pt}
		$2^{2}P_{3/2^{-}}$ & $1^{2}P_{1/2^{-}}$ & 453.24 & 4.322 & 432.87 & 81.350
		\\
		& $2^{2}P_{1/2^{-}}$ & 29.40 & 0.004 & 11.92 & 0.007
		\\
		& $2^{2}S_{1/2^{+}}$ & 231.31 & 8.566 & 219.15 & 795.667
		\\
		\hline
		\noalign{\vskip 2pt}
		$3^{2}P_{1/2^{-}}$ & $1^{2}P_{3/2^{-}}$ & 688.84 & 22.922 & 727.01 & 596.696
		\\
		& $2^{2}P_{3/2^{-}}$ & 289.99 & 4.395 & 315.93 & 137.407
		\\
		\hline
		\noalign{\vskip 2pt}
		$3^{2}P_{3/2^{-}}$ & $1^{2}P_{1/2^{-}}$ & 744.98 & 14.905 & 750.44 & 330.800
		\\
		& $2^{2}P_{1/2^{-}}$ &345.30 & 3.693 & 338.59 & 84.167
		\\
		& $3^{2}P_{1/2^{-}}$ & 29.05 & 0.004 & 11.24 & 0.006
		\\
		& $2^{2}S_{1/2^{+}}$ & 535.7 & 47.814 & 541.35 & $7.629 \times 10^{3}$
		\\
		& $3^{2}S_{1/2^{+}}$ & 201.12 & 9.233 & 194.40 & 936.842
		\\
		\hline 
	\end{tabular*}
\end{table}

\begin{table}
	\caption{$D-$ wave states radiative decays.}
	\label{tab:10}
	\centering
	\begin{tabular*}{\textwidth}{@{\extracolsep{\fill}}cccccc}
		\hline
		\noalign{\vskip 2pt}
		Initial & Final & \multicolumn{2}{c}{$\Omega_{ccc}$} & \multicolumn{2}{c}{$\Omega_{bbb}$}
		\\
		\noalign{\vskip 2pt}
		\cline{3-4}
		\cline{5-6}
		\noalign{\vskip 2pt}
		State & State & $E_{\gamma}$(MeV) & $\Gamma$(KeV) & $E_{\gamma}$(MeV) & $\Gamma$(eV) 
		\\
		\hline
		\noalign{\vskip 2pt}
		$1^{4}D_{3/2^{+}}$ & $1^{4}D_{1/2^{+}}$ & 7.18 & 0.003 & 6.81 & 0.072
		\\
		& $1^{4}D_{5/2^{+}}$ & 80.08 & 1.462 & 86.39 & 33.197
		\\
		& $1^{2}P_{1/2^{-}}$ & 241.25 & 0.016 & 213.76 & 0.262
		\\
		& $1^{2}P_{3/2^{-}}$ & 207.39 & 0.042 & 200.55 & 3.794
		\\
		\hline
		\noalign{\vskip 2pt}
		$1^{2}D_{3/2^{+}}$ & $1^{4}D_{1/2^{+}}$ & 70.79 & 0.880 & 66.28 & 36.170
		\\
		& $1^{4}D_{3/2^{+}}$ & 63.70 & 0.039 & 59.49 & 0.064
		\\
		& $1^{4}D_{5/2^{+}}$ & 142.80 & 1.879 & 145.54 & 116.466
		\\
		& $1^{2}D_{5/2^{+}}$ & 50.12 & 0.066 & 41.37 & 1.522
		\\
		& $1^{2}P_{1/2^{-}}$ & 302.00 & 15.972 & 272.39 & $1.131 \times 10^{3}$
		\\
		\hline
		\noalign{\vskip 2pt}
		$1^{4}D_{5/2^{+}}$ & $1^{2}P_{3/2^{-}}$ & 129.33 & 0.038 & 114.84 & 2.879
		\\
		\hline
		\noalign{\vskip 2pt}
		$1^{2}D_{5/2^{+}}$ & $1^{4}D_{3/2^{+}}$ & 13.71 & $8.342 \times 10^{-5}$ & 18.18 & 0.018
		\\
		& $1^{4}D_{5/2^{+}}$ & 93.58 & 1.729 & 104.46 & 1.348
		\\
		& $1^{2}P_{3/2^{-}}$ & 220.56 & 1.946 & 218.48 & 177.194
		\\
		\hline
		\noalign{\vskip 2pt}
		$1^{4}D_{7/2^{+}}$ & $1^{4}D_{5/2^{+}}$ & 95.82 & 1.992 & 97.20 & 36.699
		\\
		\hline
		\noalign{\vskip 2pt}
		$2^{4}D_{1/2^{+}}$ & $1^{4}D_{3/2^{+}}$ & 391.88 & 606.302 & 396.31 & $2.135 \times 10^{4}$
		\\
		& $1^{2}D_{3/2^{+}}$ & 332.08 & 822.366 & 338.16 & $3.891 \times 10^{4}$
		\\
		\hline
		\noalign{\vskip 2pt}
		$2^{4}D_{3/2^{+}}$ & $1^{4}D_{1/2^{+}}$ & 405.16 & 699.883 & 409.29 & $2.481 \times 10^{4}$
		\\
		& $1^{4}D_{3/2^{+}}$ & 398.51 & 34.469 & 402.66 & 166.924
		\\
		& $1^{2}D_{3/2^{+}}$ & 338.79 & 8.754 & 344.54 & 59.067
		\\
		& $1^{4}D_{5/2^{+}}$ & 472.66 & 239.623 & 486.70 & $7.845 \times 10^{3}$
		\\
		& $1^{2}D_{5/2^{+}}$ & 385.77 & 16.896 & 384.95 & $1.294 \times 10^{3}$
		\\
		& $2^{4}D_{1/2^{+}}$ & 7.13 & 0.010 & 6.52 & 0.198
		\\
		& $2^{4}D_{5/2^{+}}$ & 35.74 & 3.254 & 42.23 & 93.797
		\\
		& $2^{2}D_{5/2^{+}}$ & 5.59 & 0.002 & 3.85 & 0.081
		\\
		& $1^{2}P_{1/2^{-}}$ & 621.91 & 0.035 & 610.65 & 0.005
		\\
		& $1^{2}P_{3/2^{-}}$ & 590.56 & 0.719 & 597.80 & 110.629
		\\
		& $2^{2}P_{1/2^{-}}$ & 212.23 & 0.089 & 194.83 & 2.386
		\\
		& $2^{2}P_{3/2^{-}}$ & 183.81 & 0.228 & 183.06 & 29.872
		\\
		\hline
		\noalign{\vskip 2pt}
		$2^{2}D_{3/2^{+}}$ & $1^{4}D_{1/2^{+}}$ & 428.52 & 95.693 & 428.99 & $9.569 \times 10^{4}$
		\\
		& $1^{4}D_{3/2^{+}}$ & 421.90 & 2.266 & 422.37 & 14.185
		\\
		& $1^{4}D_{5/2^{+}}$ & 495.72 & 53.701 & 506.32 & $4.268 \times 10^{3}$
		\\
		& $1^{2}D_{5/2^{+}}$ & 409.22 & 16.494 & 404.68 & $1.058 \times 10^{3}$
		\\
		& $2^{4}D_{1/2^{+}}$ & 32.28 & 0.164 & 26.76 & 6.107
		\\
		& $2^{4}D_{3/2^{+}}$ & 25.18 & 0.001 & 20.25 & 0.002
		\\
		& $2^{4}D_{5/2^{+}}$ & 60.77 & 4.178 & 62.42 & 191.67
		\\
		& $2^{2}D_{5/2^{+}}$ & 30.76 & 0.098 & 24.09 & 9.716
		\\
		& $1^{2}P_{1/2^{-}}$ & 644.30 & 85.553 & 630.08 & $9.063 \times 10^{3}$
		\\
		& $2^{2}P_{1/2^{-}}$ & 236.46 & 46.863 & 214.81 & $3.717 \times 10^{3}$
		\\
		\hline
		\noalign{\vskip 2pt}
		$2^{4}D_{5/2^{+}}$ & $1^{4}D_{3/2^{+}}$ & 365.10 & 465.830 & 361.44 & $1.027 \times 10^{4}$
		\\
		& $1^{2}D_{3/2^{+}}$ & 304.99 & 959.259 & 303.15 & $2.366 \times 10^{4}$ 
		\\
		& $1^{2}D_{5/2^{+}}$ & 352.29 & 238.194 & 343.68 & 53.157
		\\
		& $1^{4}D_{7/2^{+}}$ & 350.16 & 793.797 & 350.82 & $1.274 \times 10^{4}$
		\\
		& $1^{2}P_{3/2^{-}}$ & 558.39 & 16.274 & 557.13 & $1.813 \times 10^{3}$
		\\
		& $2^{2}P_{3/2^{-}}$ & 149.02 & 4.026 & 141.22 & 358.489
		\\
		\hline
		\noalign{\vskip 2pt}
		$2^{2}D_{5/2^{+}}$ & $1^{4}D_{3/2^{+}}$ & 393.30 & 194.923 & 398.90 & $2.076 \times 10^{4}$
		\\
		& $1^{2}D_{3/2^{+}}$ & 333.52 & 520.284 & 340.77 & $5.525 \times 10^{4}$
		\\
		& $1^{4}D_{5/2^{+}}$ & 467.54 & 129.216 & 482.99 & 513.994
		\\
		& $1^{2}D_{5/2^{+}}$ & 380.56 & 274.015 & 381.19 & 73.709
		\\
		& $1^{4}D_{7/2^{+}}$ & 378.44 & 727.884 & 388.32 & $4.216 \times 10^{4}$
		\\
		& $2^{4}D_{5/2^{+}}$ & 30.18 & 1.134 & 38.39 & 5.928
		\\
		& $1^{2}P_{3/2^{-}}$ & 585.55 & 4.015 & 594.10 & $1.276 \times 10^{3}$
		\\
		& $2^{2}P_{3/2^{-}}$ & 178.39 & 1.601 & 179.25 & 628.480
		\\
		\hline
		\noalign{\vskip 2pt}
		$2^{4}D_{7/2^{+}}$ & $1^{4}D_{5/2^{+}}$ & 487.76 & 132.952 & 496.83 & $3.699 \times 10^{3}$
		\\
		& $1^{2}D_{5/2^{+}}$ & 401.13 & 53.122 & 395.13 & $1.633 \times 10^{3}$
		\\
		& $2^{4}D_{5/2^{+}}$ & 52.13 & 4.577 & 52.65 & 80.187
		\\
		& $2^{2}D_{5/2^{+}}$ & 22.07 & 0.035 & 14.29 & 0.978
		\\
		\hline 
	\end{tabular*}
\end{table}

\begin{table}
	\caption{$F$ and $G-$ wave states radiative decays.}
	\label{tab:11}
	\centering
	\begin{tabular*}{\textwidth}{@{\extracolsep{\fill}}cccccc}
		\hline
		\noalign{\vskip 2pt}
		Initial & Final & \multicolumn{2}{c}{$\Omega_{ccc}$} & \multicolumn{2}{c}{$\Omega_{bbb}$}
		\\
		\noalign{\vskip 2pt}
		\cline{3-4}
		\cline{5-6}
		\noalign{\vskip 2pt}
		State & State & $E_{\gamma}$(MeV) & $\Gamma$(KeV) & $E_{\gamma}$(MeV) & $\Gamma$(eV) 
		\\
		\hline
		\noalign{\vskip 2pt}
		$1^{2}F_{5/2^{-}}$ & $1^{4}D_{3/2^{+}}$ & 257.96 & 0.034 & 247.45 & 0.712
		\\
		& $1^{2}D_{3/2^{+}}$ & 196.60 & 10.001 & 188.71 & 859.245
		\\
		& $1^{2}D_{5/2^{+}}$ & 244.88 & 0.169 & 229.55 & 22.090
		\\
		\hline
		\noalign{\vskip 2pt}
		$1^{2}F_{7/2^{-}}$ & $1^{4}D_{5/2^{+}}$& 331.12 & 1.092 & 334.53 & 107.859  
		\\
		& $1^{2}D_{5/2^{+}}$ & 241.79 & 6.608 & 231.69 & 600.905
		\\
		\hline
		\noalign{\vskip 2pt}
		$2^{2}F_{5/2^{-}}$ & $1^{2}F_{7/2^{-}}$ & 330.82 & 3.176 & 336.19 & 86.381
		\\
		& $1^{4}D_{3/2^{+}}$ & 570.68 & 0.225 & 580.24 & 7.573
		\\
		& $1^{2}D_{3/2^{+}}$ & 512.90 & 155.256 & 522.81 & $2.313 \times 10^{4}$
		\\
		& $1^{2}D_{5/2^{+}}$ & 558.36 & 1.079 & 562.74 & 236.763
		\\
		& $2^{4}D_{3/2^{+}}$ & 185.14 & 0.102 & 182.41 & 3.797
		\\
		& $2^{2}D_{3/2^{+}}$ & 160.67 & 21.514 & 162.38 & $2.209 \times 10^{3}$
		\\
		& $2^{2}D_{5/2^{+}}$ & 190.56 & 0.785 & 186.21 & 131.246
		\\
		\hline
		\noalign{\vskip 2pt}
		$2^{2}F_{7/2^{-}}$ & $1^{2}F_{5/2^{-}}$ & 325.59 & 13.702 & 340.45 & 403.03
		\\
		& $1^{4}D_{5/2^{+}}$ & 640.36 & 30.383 & 665.38 & $4.969 \times 10^{3}$
		\\
		& $1^{2}D_{5/2^{+}}$ & 556.26 & 415.784 & 564.80 & $6.682 \times 10^{4}$
		\\
		& $2^{4}D_{5/2^{+}}$ & 217.49 & 89.784 & 226.28 & $1.069 \times 10^{4}$
		\\
		& $2^{2}D_{5/2^{+}}$ & 188.31 & 21.096 & 188.36 & $6.700 \times 10^{3}$
		\\
		\hline
		\noalign{\vskip 2pt}
		$1^{4}G_{7/2^{+}}$ & $1^{4}G_{5/2^{+}}$ & 12.77 & 0.863 & 20.70 & 85.592
		\\
		& $1^{2}F_{5/2^{-}}$ & 196.58 & 3.549 & 194.82 & 341.954
		\\
		& $1^{2}F_{7/2^{-}}$ & 199.69 & 1.436 & 192.66 & 142.562
		\\
		\hline
		\noalign{\vskip 2pt}
		$1^{2}G_{7/2^{+}}$ & $1^{4}G_{5/2^{+}}$ & 36.39 & 0.139 & 38.35 & 3.354
		\\
		& $1^{4}G_{7/2^{+}}$ & 23.68 & 0.037 & 17.67 & 0.393
		\\
		& $1^{2}F_{5/2^{-}}$ & 219.43 & 8.964 & 212.26 & 861.523
		\\
		\hline 
	\end{tabular*}
\end{table}

We have calculated the radiative decay widths of $\Omega_{ccc}$ and $\Omega_{bbb}$ states using the wave functions obtained from our screened potential model, considering only the electric dipole (E1) and magnetic dipole (M1) components. The results of radiative transitions are presented in Table \ref{tab:9} for the $S$ and $P-$ wave states, Table \ref{tab:10} for the $D-$ wave states, and Table \ref{tab:11} for the $F$ and $G-$ wave states. The decay widths $(\Gamma)$  are expressed in KeV for $\Omega_{ccc}$, while for $\Omega_{bbb}$ they are given in eV and the photon energies $(E_{\gamma})$ are reported in MeV. The photon energies of $\Omega_{ccc}$ and $\Omega_{bbb}$ for corresponding transitions are comparable, typically varying by no more than $10-20 \%$. This suggests that both systems share broadly similar excitation structures, despite their absolute mass scales differ by nearly a factor of three. In general, the radiative decay widths of $\Omega_{bbb}$ are often $\approx 2-4$ orders of magnitude smaller than the corresponding $\Omega_{ccc}$ widths. This suppression can be attributed to the larger bottom-quark mass, which decreases the magnetic moment and consequently diminishes the electromagnetic transition amplitudes. Despite this suppression, the hierarchical structure of dominant and subdominant decay channels is nearly equivalent in both systems, suggesting that the transition dynamics are influenced primarily by the underlying structure rather than quark flavor. For the $S-$ wave transitions in Table \ref{tab:9}, decays to the ground state $1^{4}S_{3/2^{+}}$ are suppressed for both the E1 and M1 modes, requiring higher-order multi-poles. The absence of such modes supports the oblate structure proposed in Ref. \cite{32}, and observation of these decays could provide valuable insight into the internal structure of the ground states of $\Omega_{ccc}$ and $\Omega_{bbb}$ baryons. The decay widths for the $3^{4}S_{3/2^{+}} \rightarrow 2^{2}S_{1/2^{+}}$ transition is $36.636$ KeV for $\Omega_{ccc}$ and $1.067$ KeV for $\Omega_{bbb}$, while the corresponding hyperfine transition $3^{4}S_{3/2^{+}} \rightarrow 3^{2}S_{1/2^{+}}$ exhibits a decay width suppressed by three orders of magnitude, with values of $0.055$ KeV for $\Omega_{ccc}$ and $0.046$ eV for $\Omega_{bbb}$. A similar suppression is observed for the $P-$ wave intra-shell transitions $1^{2}P_{3/2^{-}} \rightarrow 1^{2}P_{1/2^{-}}$, $2^{2}P_{3/2^{-}} \rightarrow 2^{2}P_{1/2^{-}}$ and $3^{2}P_{3/2^{-}} \rightarrow 3^{2}P_{1/2^{-}}$ with decay widths of $0.007$ KeV, $0.004$ KeV and $0.004$ KeV for $\Omega_{ccc}$, and $0.010$ eV, $0.007$ eV and $0.006$ eV for $\Omega_{bbb}$, respectively. By contrast, the inter-shell transitions $3^{2}P_{3/2^{-}} \rightarrow 1^{2}P_{1/2^{-}}$, $3^{2}P_{3/2^{-}} \rightarrow 2^{2}P_{1/2^{-}}$ and $2^{2}P_{3/2^{-}} \rightarrow 1^{2}P_{1/2^{-}}$ exhibit substantially larger decay widths. The stronger suppression of intra-shell transitions relative to inter-shell transitions in both systems reflects the relatively small spin-dependent energy splittings compared to the energy separations between principal excitation levels. Among the $P-$ wave transitions, $2^{2}P_{3/2^{-}} \rightarrow 2^{2}S_{1/2^{+}}$ is the most prominent, with decay width of $8.566$ KeV for $\Omega_{ccc}$ and $0.795$ KeV for $\Omega_{bbb}$. Comparably, the $3^{2}P_{3/2^{-}} \rightarrow 2^{2}S_{1/2^{+}}$ and $3^{2}P_{3/2^{-}} \rightarrow 3^{2}S_{1/2^{+}}$ transitions yield decay widths of $47.814$ KeV and $9.233$ KeV for $\Omega_{ccc}$ and $7.629$ KeV and $0.937$ KeV for $\Omega_{bbb}$. Transitions between states with the same parity $(P_{i} = P_{f})$ are $M1$ in nature and are consequently suppressed relative to transitions between states of opposite parity $(P_{i} = -P_{f})$, which occur through the $E1$ mechanism. This contradicts the findings of Ref. \cite{85}, where large decay widths are obtained between states having the same parity $(P_{i} = P_{f})$ and total spin $(S_{i} = S_{f})$. The radiative transitions presented in Table \ref{tab:10} for $D-$ wave states exhibits a considerably richer structure due to the larger number of accessible final states. The transitions within the same $D-$ wave multiplets, such as  $1^{4}D_{3/2^{+}} \rightarrow 1^{4}D_{1/2^{+}}, 1^{4}D_{5/2^{+}}$, $1^{2}D_{3/2^{+}} \rightarrow 1^{4}D_{1/2^{+}}, 1^{4}D_{3/2^{+}}, 1^{4}D_{5/2^{+}}$, $1^{2}D_{3/2^{+}} \rightarrow 1^{2}D_{5/2^{+}}$, $1^{2}D_{5/2^{+}} \rightarrow 1^{4}D_{3/2^{+}}, 1^{4}D_{5/2^{+}}$, $2^{4}D_{3/2^{+}} \rightarrow 2^{4}D_{1/2^{+}}, 2^{4}D_{5/2^{+}}$, $2^{4}D_{3/2^{+}} \rightarrow 2^{2}D_{5/2^{+}}$, $2^{2}D_{3/2^{+}} \rightarrow 2^{4}D_{1/2^{+}}, 2^{4}D_{3/2^{+}}, 2^{4}D_{5/2^{+}}$, etc, have small decay widths. These intra-shell transitions are a direct consequence of the very small fine-structure splittings within a given excitation band in triply heavy baryons, reflecting the weak spin-dependent interactions characteristic of heavy-quark systems. In contrast, the $2D \rightarrow 1D$ transitions constitute the largest partial widths among the $D-$ wave states, approximately an order of magnitude above the next-strongest channels, indicating that they are strongly favored. In particular, the transitions $2^{4}D_{1/2^{+}} \rightarrow 1^{2}D_{3/2^{+}}$, $2^{4}D_{3/2^{+}} \rightarrow 1^{4}D_{1/2^{+}}$, $2^{2}D_{3/2^{+}} \rightarrow 1^{4}D_{1/2^{+}}$, $2^{4}D_{5/2^{+}} \rightarrow 1^{2}D_{3/2^{+}}$, $2^{4}D_{7/2^{+}} \rightarrow 1^{4}D_{5/2^{+}}$ constitute the dominant decay channels for their respective initial states, with predicted widths of $822.366$ KeV, $699.883$ KeV, $95.693$ KeV, $959.259$ KeV, $132.952$ KeV for $\Omega_{ccc}$, and $38.91$ KeV, $24.81$ KeV, $95.69$ KeV, $23.66$ KeV, $3.699$ KeV for $\Omega_{bbb}$, respectively. In the case of $2^{2}D_{5/2^{+}}$, the dominant decay channel in $\Omega_{ccc}$ is $2^{2}D_{5/2^{+}} \rightarrow 1^{4}D_{7/2^{+}}$ with width of $727.884$ KeV, whereas for $\Omega_{bbb}$ it is $2^{2}D_{5/2^{+}} \rightarrow 1^{2}D_{3/2^{+}}$ with width of $55.25$ KeV. For transitions from $D-$ wave states to $P-$ wave states, a clear selectivity is observed between the $n^{4}D$ and $n^{2}D$ states. In general, transitions originating from $n^{4}D$ states, such as $1^{4}D_{3/2^{+}} \rightarrow 1^{2}P_{1/2^{-}}$, $1^{4}D_{3/2^{+}} \rightarrow 1^{2}P_{3/2^{-}}$, $1^{4}D_{5/2^{+}} \rightarrow 1^{2}P_{3/2^{-}}$, $2^{4}D_{3/2^{+}} \rightarrow 1^{2}P_{1/2^{-}},1^{2}P_{3/2^{-}}$, $2^{4}D_{3/2^{+}} \rightarrow 2^{2}P_{1/2^{-}},2^{2}P_{3/2^{-}}$, $2^{4}D_{5/2^{+}} \rightarrow 2^{2}P_{3/2^{-}}$, etc are noticeably suppressed. By comparison, transitions from the corresponding $n^{2}D$ states, including $1^{2}D_{3/2^{+}} \rightarrow 1^{2}P_{1/2^{-}}$, $1^{2}D_{5/2^{+}} \rightarrow 1^{2}P_{3/2^{-}}$, $2^{2}D_{3/2^{+}} \rightarrow 1^{2}P_{1/2^{-}}, 2^{2}P_{1/2^{-}}$, $2^{2}D_{5/2^{+}} \rightarrow 1^{2}P_{3/2^{-}}, 2^{2}P_{3/2^{-}}$, etc, exhibit much larger decay widths. The notable exception is the transition $2^{4}D_{5/2^{+}} \rightarrow 1^{2}P_{3/2^{-}}$ , which remains comparatively strong despite originating from an $n^{4}D$ state. The $F$ and $G-$ wave radiative transitions are listed in Table \ref{tab:11}. Similar to the $D-$ wave states, the $F-$ wave states exhibits a clear selectivity in decay channels, where transitions to spin-doublet $^{2}D$ states are strongly favored over those to spin-quartet $^{4}D$ states. The results suggest that $F$-wave baryons predominantly decay radiative transitions to lower-lying $D$-wave states. Transition widths within the same $G$-wave multiplet are strongly suppressed owing to the very small mass splittings among the states. In contrast, decays into lower-lying $F-$ wave states exhibit substantially larger widths.

\section{\label{sec:Conclusion} Conclusion}

Using our screened potential model in the quark–diquark picture, we have analyzed the low–lying as well as radially and orbitally excited spectra of the $\Omega_{ccc}$ and $\Omega_{bbb}$ baryons and computed their leading E1/M1 radiative decay widths. Our predicted ground–state masses, $M(\Omega_{ccc}) = 4.660$ Gev and $M(\Omega_{ccc}) = 14.200$ GeV, are consistent with other model estimates. The radiative decay widths of $\Omega_{bbb}$ are suppressed by roughly three orders of magnitude relative to those of $\Omega_{ccc}$, a direct consequence of the smaller magnetic moment of the $b$ quark. Direct E1/M1 transition to the $1^{4}S_{3/2^{+}}$ ground state are suppressed, implying that higher multi-pole processes could provide valuable insight into the internal structure and charge distribution of these baryons. A detailed analysis of the radiative decay patterns has been performed, highlighting both the qualitative and quantitative differences between the $\Omega_{ccc}$ and $\Omega_{bbb}$ baryons,  as well as the systematic trends that emerge across different radial and orbital excitations. Our results provide a comprehensive and coherent picture of the spectra and radiative properties of triply heavy baryons, offering a useful reference for future experimental searches.

\section*{Acknowledgements}

One of the authors (CAB) is grateful to Manipal Academy of Higher Education (MAHE) for the financial support under the Dr. T.M.A. Pai Scholarship Program. CAB would also like to express his sincere gratitude to Kamil Serafin for valuable suggestions.

\nocite{*}

\bibliography{cccandbbbBaryonsArxiv}

\end{document}